\documentclass[a4paper,fleqn,usenatbib]{mnras}
\usepackage{newtxtext,newtxmath}
\usepackage[T1]{fontenc}
\usepackage{ae,aecompl}

\usepackage{graphicx}	
\usepackage{amsmath}	
\usepackage{amssymb}	
\usepackage{ulem}
\usepackage{hyperref}

\title[Segmenting the Universe]{Segmenting the Universe into dynamically coherent basins}

\author[Dupuy et al.]
{Alexandra Dupuy$^{1}$\thanks{E-mail: dupuy@ipnl.in2p3.fr},
H\'el\`ene M. Courtois$^{1}$,
Noam I. Libeskind$^{1,2}$,
Daniel Guinet$^{1}$
\\
$^{1}$University of Lyon, UCB Lyon 1, CNRS/IN2P3, IUF, IP2I Lyon, France\\
$^{2}$Leibniz-Institut f\"ur Astrophysik Potsdam (AIP), An der Sternwarte 16, D-14482 Potsdam, Germany}

\begin{document}

\date{Accepted....... ;}

\pagerange{\pageref{firstpage}--\pageref{lastpage}} \pubyear{2019}

\maketitle

\label{firstpage}

\begin{abstract}
 This article explores in depth a watershed concept to partition the universe, introduced in \cite{2019MNRAS.489L...1D} and applied to the {\it Cosmicflows-3} observational dataset. We present a series of tests conducted with cosmological dark matter simulations. In particular we are interested in quantifying the evolution with redshift of large scale structures when defined as segmented basins of attraction. This new dynamical definition in the field of measuring standard rulers demonstrates robustness since all basins show a density contrast $\delta$ above one (mean universe density) independently of the simulation spatial resolution or the redshift. Another major finding is that density profiles of the basins show universality in slope. Consequently, there is a unique definition of what is a gravitational watershed at large scale, that can be further used as a probe for cosmology studies. 
\end{abstract}

\begin{keywords}
large-scale structure of Universe
\end{keywords}

\section{Introduction}\label{intro} 

One of the most basic tenets of modern astronomy is  the Cosmological principle, namely that the Universe is isotropic and homogeneous on large enough scales. The galaxy distribution is uniform in all directions above the scale of inhomogeneity. Below this scale, matter and galaxies are concentrated into clustered regions connected by filaments, while some locations are empty with very few galaxies. This filamentary structure forms a complex network defined as the {\it cosmic web} \citep{1996Natur.380..603B}. With time, these large-scale structures gradually evolve into more compact structures. The clustered regions and filamentary structure of the Universe becomes more and more emphasized, while voids are drained.  

The  Cosmic Web is one of the most intriguing and striking patterns found in nature, rendering its analysis and characterization 
far from trivial. This is evidenced by the many elaborate descriptions in the literature. Some of these are based on observational data, namely redshift surveys \citep{Aragon-Calvo:2007aa, Aragon-Calvo:2010aa,2014MNRAS.440L..46A,Sousbie:2008aa,2010MNRAS.407.1449G, Sousbie:2011aa,2012ApJ...754..126F,2015MNRAS.450.3239F,2014MNRAS.438..177A,2014MNRAS.438.3465T,2015MNRAS.452.1643R,2016A&C....16...17T,Leclercq:2017aa}. Others make use numerical simulations where the full 6-dimensional matter phase space is known \citep{Hahn:2007aa,Forero-Romero:2009aa,Hoffman:2012aa,2012MNRAS.425.2443K,Cautun:2013aa}. According to these schemes the  large-scale structures is often classified as knots, filaments, sheets and voids.  All of these various methodologies, and others, are reviewed and compared in \cite{Libeskind:2018aa}. We note that unlike how the virial theorem defines a dark matter halo, there is no theory \citep[beyond][]{1970Ap......6..320D} that unambiguously defines large scale structures. Therefore, these methods do not give a unique definition of what a large scale structure is. Additionally since large scale structure is a multi-scale phenomenon, most of these measurements depend on the scale (or resolution) of the problem.

On another hand, the term {\it supercluster} is often used to define as an overall entity, the ensemble of knots, filaments, sheets and voids that form a recognizable structure \citep{1985AJ.....90.2445G,1987ang..book.....T,1988ApJ...326...19L,1989Sci...246..897G,2005ApJ...624..463G,2016A&A...588L...4L}. More specifically, superclusters are not gravitationally bound in a Universe which is expanding. However as this term is not rigorously and physically defined it has also been applied to structures that are at rest in the cosmic microwave background.  In this sense, they can be defined as basins of gravitational attraction - watersheds - at rest. The term of {\it watershed} has been introduced in cosmology, for example by \cite{2007MNRAS.380..551P}, to partition a density field into cosmic watersheds. Looking at an instantaneous moment of the Universe (today), \cite{Tully:2014aa} introduced a new kinematical definition of superclusters: basins of attraction \citep[see also][]{2015ApJ...812...17P}. We note that this is the first definition of {\it superclusters} based on dynamics (and not, for example, on arbitrary metrics). In addition to this new definition, one can identify volumes that are dual to the basins of attraction: the basins of repulsion \citep{2017NatAs...1E..36H,2017ApJ...847L...6C}. 

Pursuing the development of this concept, \cite{2019MNRAS.489L...1D} proposed a new method to automatically partition the universe into gravitational basins, by using a peculiar velocity field and {\it streamlines}. This new methodology has been applied to a velocity field reconstructed from the {\it Cosmicflows-3} catalog of observed peculiar velocities \citep{2019MNRAS.488.5438G}. The success of \cite{2019MNRAS.489L...1D} was the ability to apply a theoretical concept to a reconstructed universe, thereby identifying real structures via their signature in the peculiar velocity field. Moreover, the new technique introduced in this paper allowed to confirm kinematically the hidden Vela supercluster, discovered by \cite{2017MNRAS.466L..29K}, from the {\it Cosmicflows-3} data \citep{2019MNRAS.490L..57C}.

Gravitational watersheds result from the intertwined effects of gravitation and expansion playing on very large scales. The growth rate of such large scales structures with time, is a direct measurement of the cosmology. These are also the largest structures  known in the Universe. Studying their size and shape can give clues on how the total (baryonic and dark) mass is distributed, and consequently help discriminate between Dark Matter (DM) models since for example WDM clusters less than CDM. It is also yet to be discovered if there are scale free relations, induced by astrophysical processes, in sizes and density profiles of such large scale structures.

In this paper, the segmentation methodology introduced previously in \cite{2019MNRAS.489L...1D} is applied to a $\Lambda$CDM dark matter simulation in order to conduct a series of tests. More specifically, the main goal is to study the cosmic evolution of large-scale structures when they are defined as gravitational basins. The partitioning algorithm is tested at various scales and various redshifts in order to quantify the evolution with time and scale of various properties of basins, such as their mass or mean density. 

 The paper is organized as follows. After this short introduction, the simulation considered throughout this paper, as well as the segmentation algorithm, are described in Section \ref{sec:method}. Section \ref{sec:results} presents the results and is organized as follows. After testing the various algorithmic parameters on $z=0$ basins of attraction in Section \ref{sec:results_z0}, we move in Section \ref{sec:results_allz} onto a discussion covering the time evolution of basins of attraction, with analyses of their total number, respective volumes and masses. The paper ends with a short conclusion in Section \ref{sec:conclusion}.

\section{Methodology}\label{sec:method} 

In this section, the newly developed segmentation methodology, and the cosmological numerical simulations used throughout this study, are briefly described. 

\subsection{Simulations}\label{sec:simulations}

For this article, the ``Small Multidark'' simulation (SMD) is used. SMD is part of the Multidark suite of $N$-body simulations\footnote{ see https://www.cosmosim.org for more detailed information.} \citep{2016MNRAS.457.4340K}.
This is a dark-matter-only $N$-body simulation of 3840$^3$ particles that assumes a $\Lambda$CDM power spectrum of fluctuations according to the Planck cosmological  parameters (\citealt{2016MNRAS.462..893R};  $\Omega_{\Lambda} = 0.69$, $\Omega_{M} = 0.33$, $\sigma_{8} = 0.83$ and $H_{0} = 67.77$ km s$^{-1}$ Mpc$^{-1}$). The simulation is carried out in a box of side length $L_{\rm box} = 400$ $h^{-1}$Mpc and thus  achieves a mass resolution of $9.4 \times 10^7$ M$_\odot/h$ per particle and a spatial softening length of $1.5h^{-1}$kpc.  

 A clouds-in-cells (CIC) algorithm has been applied to the SMD simulation in order to derive the density and velocity fields in a $256^3$ grid, from the distribution of the particles in the simulation box, resulting in a spatial resolution of $1.5625$ Mpc/$h$. For the tests presented in this article, a final Gaussian smooth is applied to the velocity field, with various smoothing lengths $r_s = 1.5$ Mpc/$h$, $3$ Mpc/$h$, $5$ Mpc/$h$, $7.5$ Mpc/$h$, $10$ Mpc/$h$, $12.5$ Mpc/$h$ and $15$ Mpc/$h$.

\subsection{Segmenting the Universe}\label{sec:segmentation}

The methodology considered throughout this paper to partition the simulated Universe into gravitational basins makes use of streamlines - i.e. computed integration curves which, in the case of a time-independent linear velocity field, are the paths tangent to the local value of the velocity field. The three main steps of this algorithm are: computing streamlines, identifying locations of attractors (or repellers), deriving their corresponding watersheds, ie basins of attraction (or repulsion). The reader may refer to \citep{2019MNRAS.489L...1D} for a more detailed description.

 The first and main step of this method is to generate streamlines starting from every voxel of the velocity grid. One can generate a streamline $\vec{s}$, from a given seed point $\vec{s}_0$, by integrating spatially the components of the velocity field $\vec{v}$. The integration is carried out by the fourth order Runge-Kutta numerical integrator (RK4) after setting two parameters: the maximum number of integrations, which defines the length of streamlines $l_s$, and the the integration step $\Delta\tau$. The direction of integration, i.e integrating $\vec{v}$ or $-1\times\vec{v}$, allows to visualize basins of attraction (BOA), or basins of repulsion (BOR) respectively. 

 Secondly, the {\it center} of the basins - the locations where streamlines converge/end, the critical points of the velocity field - need to be identified. Such positions can be detected by looking only at the end-point of the streamlines, i.e the positions at which streamlines have been terminated. Namely, the number of terminated streamlines is computed for each voxel. Attractors (or repellers) can be easily identified as they correspond to the local maxima of this value. In order to find such local maxima, a simple approach is implement in the algorithm: the grid is scanned voxel by voxel, and if the value of a given voxel is greater than the 26 voxels that are adjacent to it (namely than those that share either a face, an edge or a corner), then it is considered a local maximum. In order to scan the grid faster and skip voxels that are less likely to correspond to local maxima, one can add a threshold $n_{s,\mathrm{max}}$ on the number of streamlines ending in a voxel, i.e while the 3D grid is scanned, only the voxels that have a value greater than this threshold are checked. This is equivalent to setting a threshold on the volume and size of the basins.

 Lastly, the gravitational basins are built by linking the seed-points voxels of the streamlines to their respective single end-point voxel (local maxima - attractors or repellers). As, in our case, each voxel represents the seed point of a streamline ending at one of the local maxima locations, each voxel that is not a local maxima can be allocated to the basin in which its streamline stops.

\section{Results}\label{sec:results}

This section is organized as follows. First, the parameters of the algorithm described in the previous section, are tested on the present-day ($z=0$) simulated velocity field. Afterwards, the segmentation methodology is applied to various redshifts of the SMD simulation, in order to study the expected cosmic evolution of gravitational basins.

\subsection{Properties of $z=0$ basins }\label{sec:results_z0}

In this section we examine how the properties of $z=0$ basins are affected by algorithmic choices.  A fiducial Gaussian smoothing of length 1.50 Mpc/$h$ is applied.  

\subsubsection{Algorithmic parameters: $l_s$, $\Delta\tau$, $n_{s, \rm max}$}

As stated in section \ref{sec:segmentation}, the computation of streamlines requires to set two parameters: the maximum streamlines' length $l_s$ and the integration step $\Delta\tau$. 

\begin{figure*}
\includegraphics[width=\linewidth]{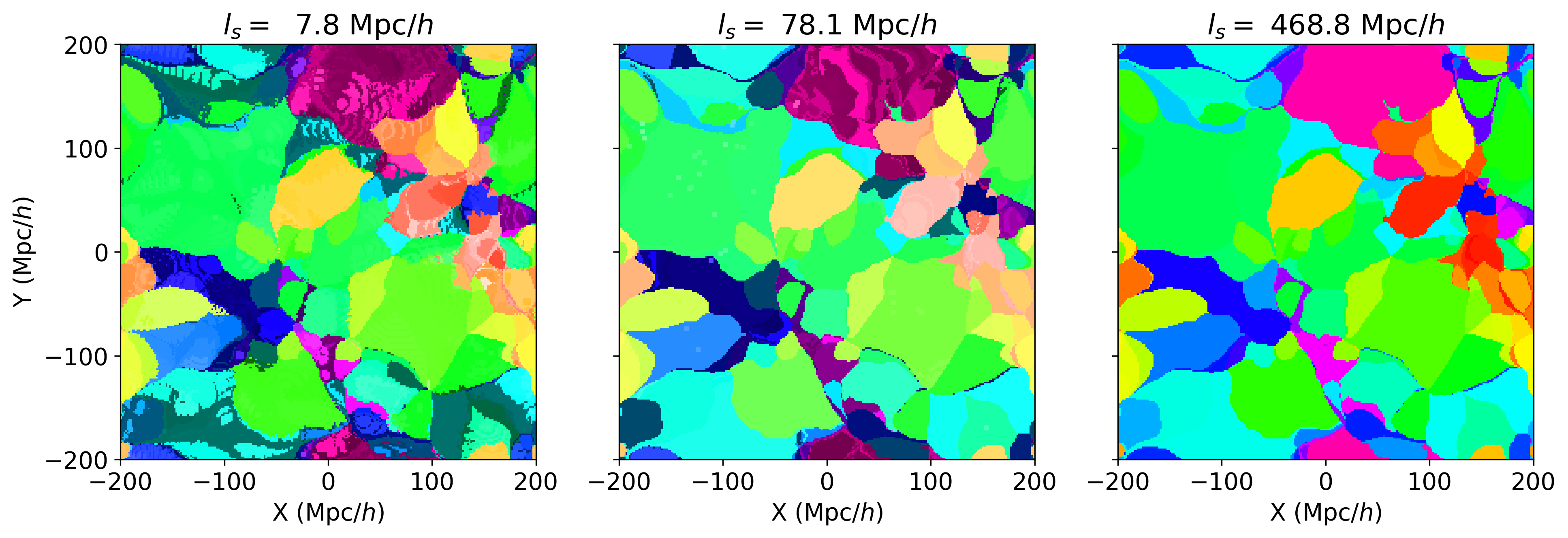}
\caption{Change of gravitational basins when varying the maximum allowed length of streamlines $l_s$. All three panels represent XY slices in Supergalactic Cartesian coordinates, centered on SGZ$=0$ Mpc/$h$ of the Small Multidark Simulation at z=0. Each colored enclosed surface corresponds to the 2D projection of a basin of attraction identified automatically by our segmentation methodology. Three different maximum length of streamlines are considered from left to right : $l_s = 7.8$ Mpc/$h$, $78.1$ Mpc/$h$ and $468.8$ Mpc/$h$.}
\label{fig:length}
\end{figure*}

Figure \ref{fig:length} shows the same three XY slices centered on Z = 0 Mpc/$h$. The three panels display the segmentation of the SMD present-day velocity field, where the maximum streamlines' length has been set to three different values: $l_s = 7.8$ Mpc/$h$, $78.1$ Mpc/$h$ and $468.8$ Mpc/$h$ respectively. A total of 10 different values for $l_s$ have been tested, but for the sake of clarty, only 3 of them are shown here. Note that the largest length tested here is higher than the size of the grid ($400$ Mpc/$h$), thus it corresponds to the largest length we should test. Each coloured enclosed region represents a 2D slice through a single basin of attraction. On the left panel, some {\it fingerprint}-like patterns can be observed, while they tend to disappear as the streamlines length are allowed to be longer, as one can see in middle and right panels. Such patterns are artificial and are due to the abrupt cessation of streamline length - their existence suggests that the stream line length limit is too small. Some  additional flaws can still be seen in the middle panel. For example, some tiny basins can be identified inside the larger basin located, for example, at X$=-150$ Mpc/$h$ and Y$=50$ Mpc/$h$. These small basins can no longer be seen when increasing more the allowed maximum length of streamlines, as seen on the right panel. 

\begin{figure*}
\includegraphics[width=\linewidth]{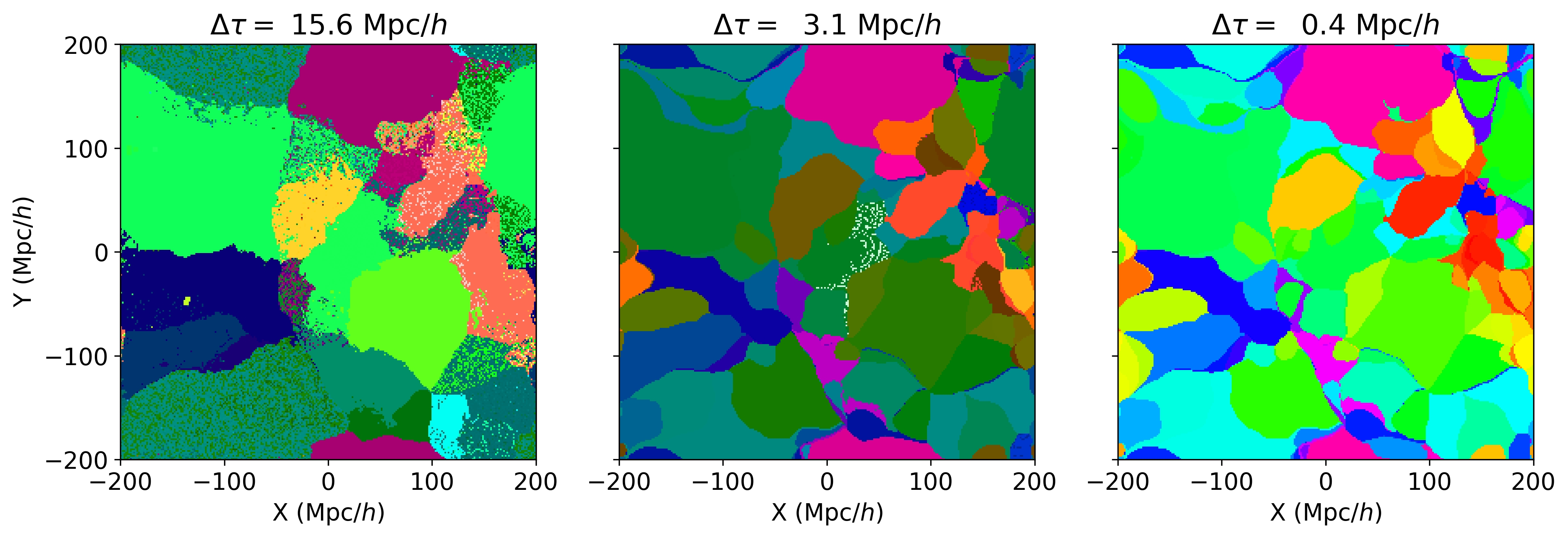}
\caption{Variations in watersheds partitions according to the computationally preset streamlines integration step $\Delta\tau$. All three panels represent XY slices in Supergalactic Cartesian coordinates, centered on SGZ$=0$ Mpc/$h$. Each colored filled isocontour (imprint, surface) corresponds to the 2D slice thru a basin of attraction identified by our segmentation methodology. Three different Runge-Kutta integration steps are considered: $\Delta\tau = 15.6$ Mpc/$h$, $3.1$ Mpc/$h$, $0.4$ Mpc/$h$ from left to right.}
\label{fig:step}
\end{figure*}

 Figure \ref{fig:step} displays the segmentation of n velocity field on three X-Y slices centered on Z$=0$ Mpc/$h$. In this test, the integration step $\Delta\tau$ is modified: from left to right $\Delta\tau = 15.6$ Mpc/$h$, $3.1$ Mpc/$h$, $0.4$ Mpc/$h$. On the left panel, the large integration step leads to grainy patterns within basins, and the inability to identify small-scale basin. The middle and right panels show that the grainy patterns vanish when decreasing $\Delta\tau$. In the middle panel, such patterns can still be observed for example at the location $(\mathrm{X},\mathrm{Y}) = (50,25)$ Mpc/$h$, $(150,50)$ Mpc/$h$ and $(190,-190)$ Mpc/$h$. However these patterns are not visible anymore in the right panel. Moreover, the smallest basins that were not identified in the left and middle panels are now detected in the right panel: for example around $(\mathrm{X},\mathrm{Y}) = $  $(-50,150)$ Mpc/$h$ in yellow,  $(50,-150)$ Mpc/$h$ in purple and  at $(-175,-100)$ Mpc/$h$ in green and yellow.

\begin{figure*}
\includegraphics[width=\linewidth]{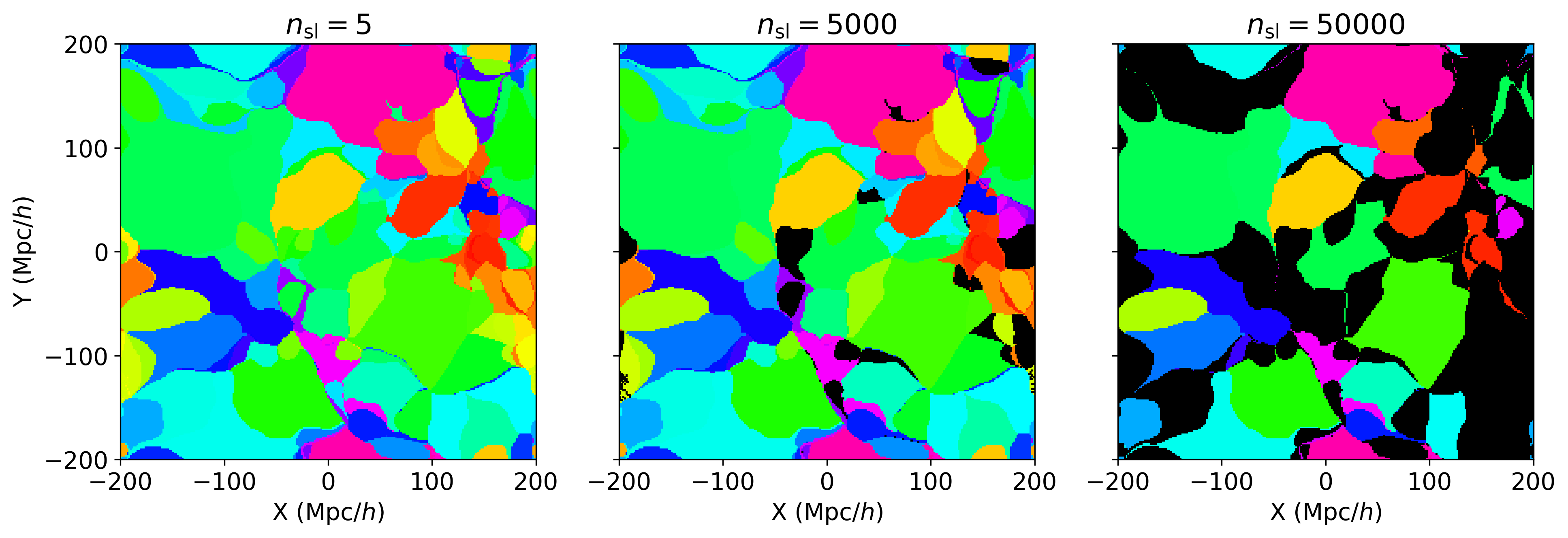}
\caption{Test of the computational pre-set parameter corresponding to the threshold in the maximum number of streamlines required in order to define a basin. From left to right: 5, 5,000 and 50,000. For reference this is Snapshot 088 of the Small Multidark Simulation with an applied spatial gaussian smoothing of 1.50 Mpc/h. The right panel shows that larger watersheds are identified when looking for the highest concentrations of streamlines in space.}
\label{fig:threshold}
\end{figure*}

 Figure \ref{fig:threshold} presents how the segmentation of the z=0 SMD velocity field is affected by the choice of a pre-set parameter corresponding to the threshold in the maximum number of streamlines ending in a voxel $n_{s,\mathrm{max}}$ required to define a gravitational basin. On each panel, a X-Y slice centered on Z$=0$ Mpc/$h$ is displayed. Similarly to Figures \ref{fig:length} and \ref{fig:step}, the colored regions represent the 2D projection of segmented basins of attraction. Voxels that do not belong to any basin of attraction are colored in black. From left to right, different values of $n_{s,\mathrm{max}}$ are considered when looking for local maxima : $n_{s,\mathrm{max}} = 5$, $5,000$ and $50,000$ streamlines. One can see, on the left panel, that a low threshold allows to identify all basins of all scales. However, the middle and right panels show that increasing the threshold prevents the algorithm to detect the small-scale basins: 3\% and 42\% of the grid is not segmented when $n_{s,\mathrm{max}} = 5,000$ (middle) and $50,000$ (right) respectively.

\subsubsection{Parameter selection}
\begin{figure*}
\includegraphics[width=\linewidth]{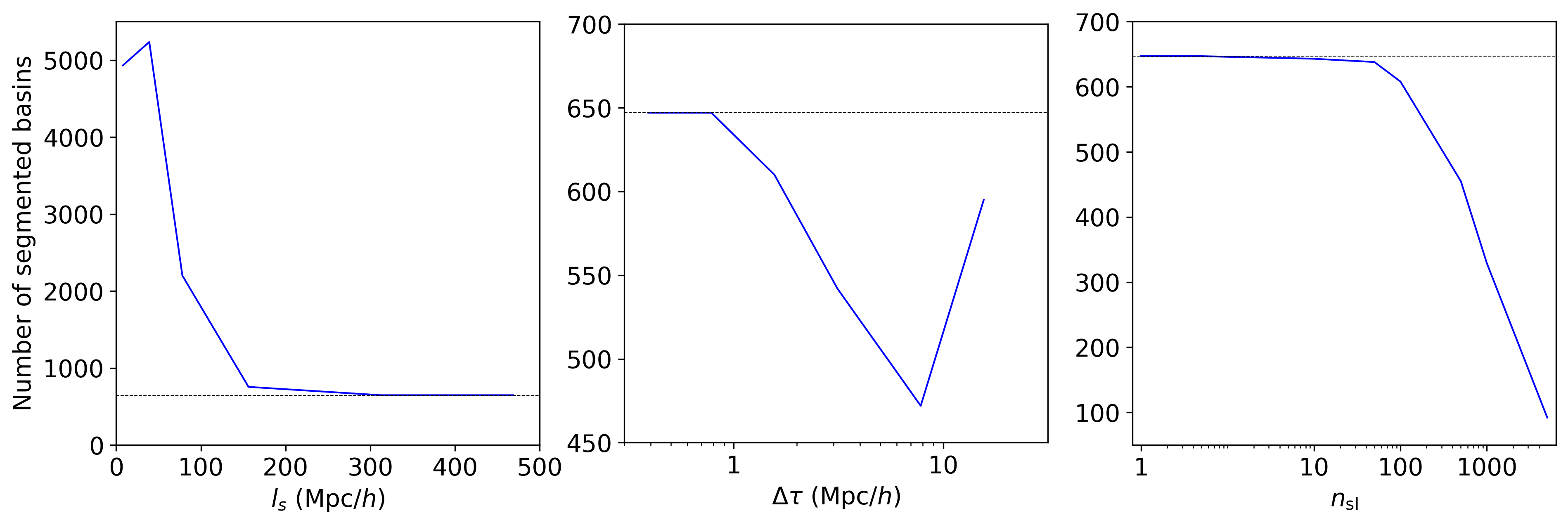}
\caption{From left to right: number of basins as a function of the maximum streamline length, Runge-Kutta integration step and threshold in number of streamlines to define a watershed. In all plots the number of converge to 647 at the following values: $l_s = 312.5$ Mpc/$h$, $\Delta\tau = 0.8$ Mpc/$h$ and $n_{s,\rm max} = 5$. These values are the ones used for the next section analysis with varying redshifts. For reference these plots are made with the SMD velocity field at $z=0$ and with an applied spatial Gaussian smoothing of 1.50 Mpc/$h$.}
\label{fig:convergence}
\end{figure*}

 We saw above how the space segmentation obtained from the SMD velocity field varies according to algorithmically preset values. However, the panels of Figure \ref{fig:convergence} show the evolution of the number of basins $N_b$ identified by the segmentation method as a function of these three parameters: the maximum length of streamlines $l_s$, the Runge-Kutta integration step $\Delta\tau$ and the threshold in number of streamlines $n_{s,\mathrm{max}}$, from left to right respectively. The number of basins converge to $N_b = 647$ from $l_s = 312.5$ Mpc/$h$, $\Delta\tau = 0.8$ Mpc/$h$ and $n_{s,\rm max} = 5$.
This convergence allows to pursue tests  of the possible watersheds evolution with redshift in Section \ref{sec:conclusion}.

\subsubsection{Smoothing scale}\label{sec:smoothingscale}

The gravitational basins identified in a given velocity field may also depend on parameters that are not directly related to the numerics of the segmentation algorithm described above in Section \ref{sec:method}. For example, the partition of a given velocity field also depends on the scale of the Gaussian smoothing of the velocity field. This is a physical choice, not a numerical one.

\begin{figure*}
\includegraphics[width=\linewidth]{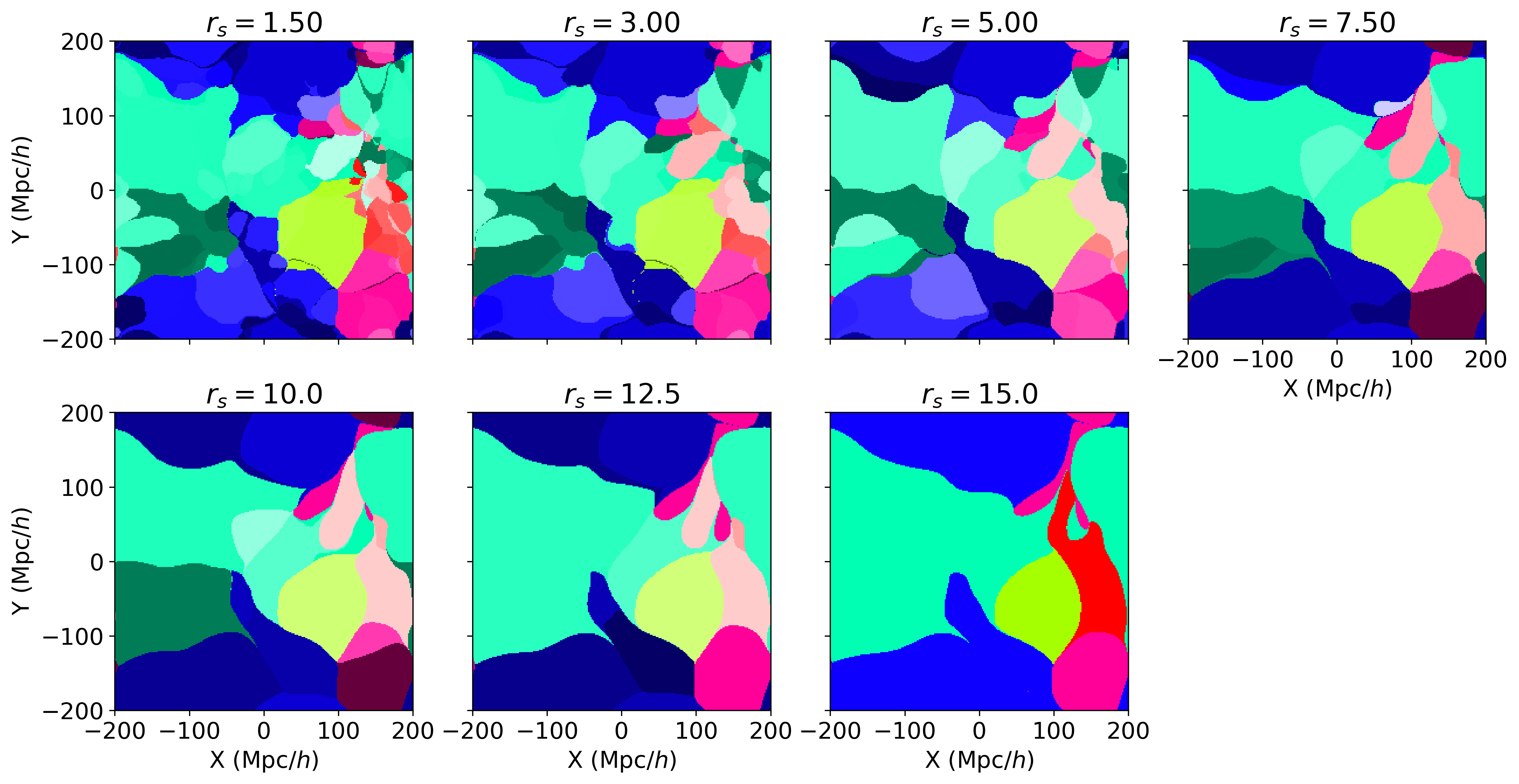}
\caption{Gravitational basins and smoothing scale $r_s$. The seven panels display the same SGX-SGY slice centered on SGZ$=0$ Mpc/$h$. Each colored (expect black) filled isocontour correspond to a single basin of attraction identified by our segmentation code. The segmentation algorithm has been applied to the SMD velocity field smoothed with six different $r_s$: $r_s = 1.50$ Mpc/$h$, $3$ Mpc/$h$, $5$ Mpc/$h$, $7.50$ Mpc/$h$, $10$ Mpc/$h$, $12.5$ Mpc/$h$ and $15$ Mpc/$h$ from left to right and up to down.}
\label{fig:smoothing}
\end{figure*}

 Figure \ref{fig:smoothing} shows how the segmentation evolves with the smoothing scale applied on the SMD velocity field: $r_s = 1.50$ Mpc/$h$, $3$ Mpc/$h$, $5$ Mpc/$h$, $7.50$ Mpc/$h$, $10$ Mpc/$h$, $12.5$ Mpc/$h$ and $15$ Mpc/$h$ from left to right, and up to down. We note that the size of a CIC grid cell sets the smallest smoothing we may test, 1.5Mpc$/h$. In  principle the largest smoothing scale is set by the size of the box, but would be trivial to test.
 
 The three panels display the same X-Y slice centered on Z$=0$ Mpc/$h$. Each colored (expect black) filled region correspond to a single basin of attraction identified by our segmentation code. As expected, it is clear that a larger smoothing scale leads to larger gravitational basins, while a small $r_s$ allows the identification of small-scale features.
 
We note that unlike convergence tests (e.g. Fig.~\ref{fig:convergence}) which can be used to pick algorthmic parameters, the adopted smoothing scale on which basins are computed must be physically motivated and remains dependent on the question one wishes to ask.

\subsection{Cosmic evolution of basins}\label{sec:results_allz}

This section is dedicated to the study of the cosmic evolution of gravitational basins. Snapshots representing the SMD velocity field at different redshifts $z$ have been selected in order to analyze how the basins evolve with time: $z = 2.89$, $2.48$, $2.14$, $1.44$, $1.00$, $0.74$, $0.51$, $0.40$, $0.29$, $0.20$, $0$. The seven smoothing scales $r_s$ considered above in Section \ref{sec:smoothingscale} are considered throughout this section. 

\begin{figure*}
\includegraphics[width=0.7\linewidth]{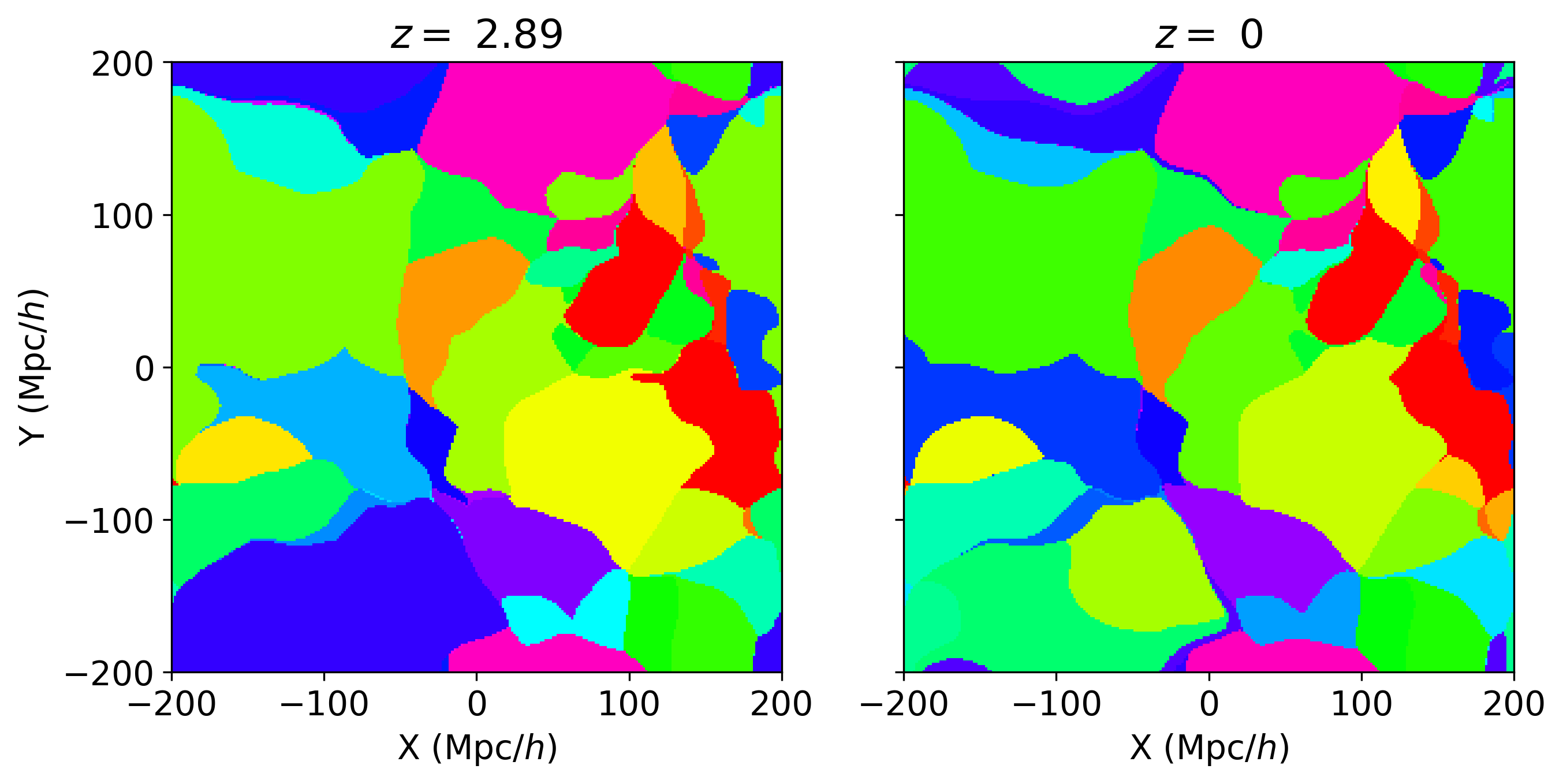}
\caption{Cosmic evolution of basins of attraction. The right and left panels represent the same X-Y slice centered at Z=0 Mpc/$h$, at redshifts $z=2.98$ and $z=0$ respectively. As usual, each  colored enclosed surface corresponds to the 2D projection of a basin of attraction identified automatically by our segmentation algorithm.}
\label{fig:redshift}
\end{figure*}

The segmentation obtained for the SMD velocity field for two different redshifts is displayed in Figure \ref{fig:redshift}. The left panel corresponds to $z = 2.89$ while the right panel corresponds to $z = 0$. In both cases the velocity field is smoothed with a scale $r_s = 5$ Mpc/$h$. The segmentation of both velocity fields is very similar. The shapes of the basins are nearly identical (see for example the pink basin located at (X,Y)=(50,150) Mpc/$h$). The fact that structures become more contrasted with time may explain the small differences. Despite the fact that the density and velocity fields are less contrasted at $z=2.89$ than at $z=0$, the similarity in the appearance of the basins at these two redshifts is an indication that the smoothing choosen, 5Mpc$/h$ is well above the scale of non-linearity  \citep[e.g. see][]{2014MNRAS.441.1974L} and thus linearizes the fields.

\begin{figure}
\includegraphics[width=\linewidth]{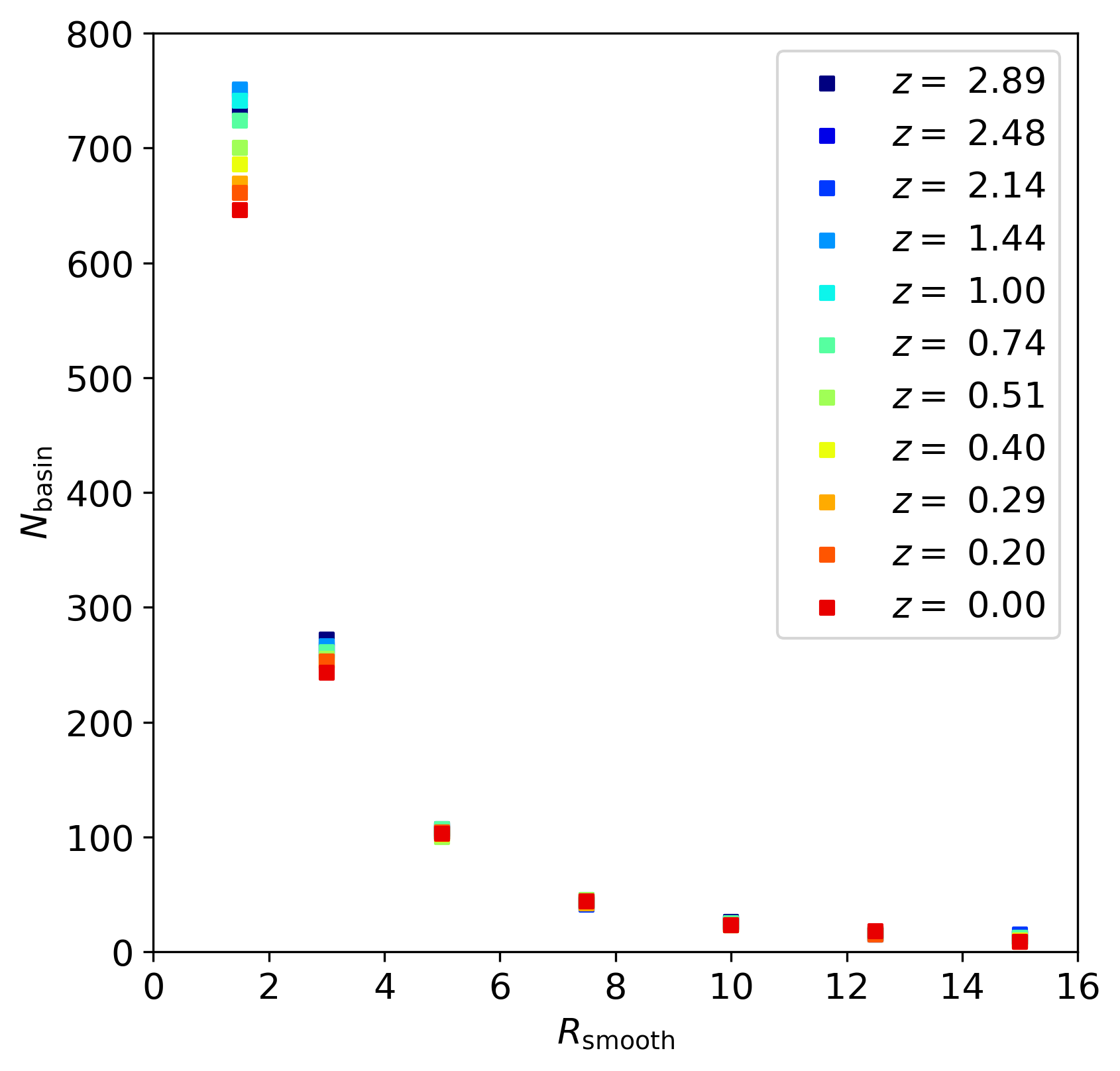}
\caption{Number of basins of attraction identified in the SMD velocity field as a function of the smoothing scale $r_s$. The scattered points are colored depending the the redshift considered from $z=2.89$ in blue to $z=0$ in red. The shapes of the basins do not evolve with time.}
\label{fig:znumber}
\end{figure}

 Figure \ref{fig:znumber} shows the number of basins of attraction identified in the simulation as a function of the smoothing scale $r_s$. The scattered points are colored depending the the redshift considered from $z=2.89$ in blue to $z=0$ in red. First, this figure shows that the number of basins in a given velocity field decreases when the smoothing scale $r_s$ increases, irrespective of redshift. This is expected as we saw above in section \ref{sec:smoothingscale} that the gravitational basins get larger as the smoothing scale increases (also see Figure \ref{fig:smoothing}). Secondly, one may notice very little dependence in redshift. The largest scatter between the different redshifts can be observed for the smallest $r_s$, however this scatter disappears as $r_s$ increases. This means that approximately the same number of basins are identified for the same $r_s$, regardless of the redshift. Again this is not unexpected. Smoothings that are large enough such that the rms of the normalized density field is greater unity (i.e. $\rho/\rho_{\rm mean} > 1$) will result in identical basins, albeit of different density contrasts.

\begin{figure}
\includegraphics[width=\linewidth]{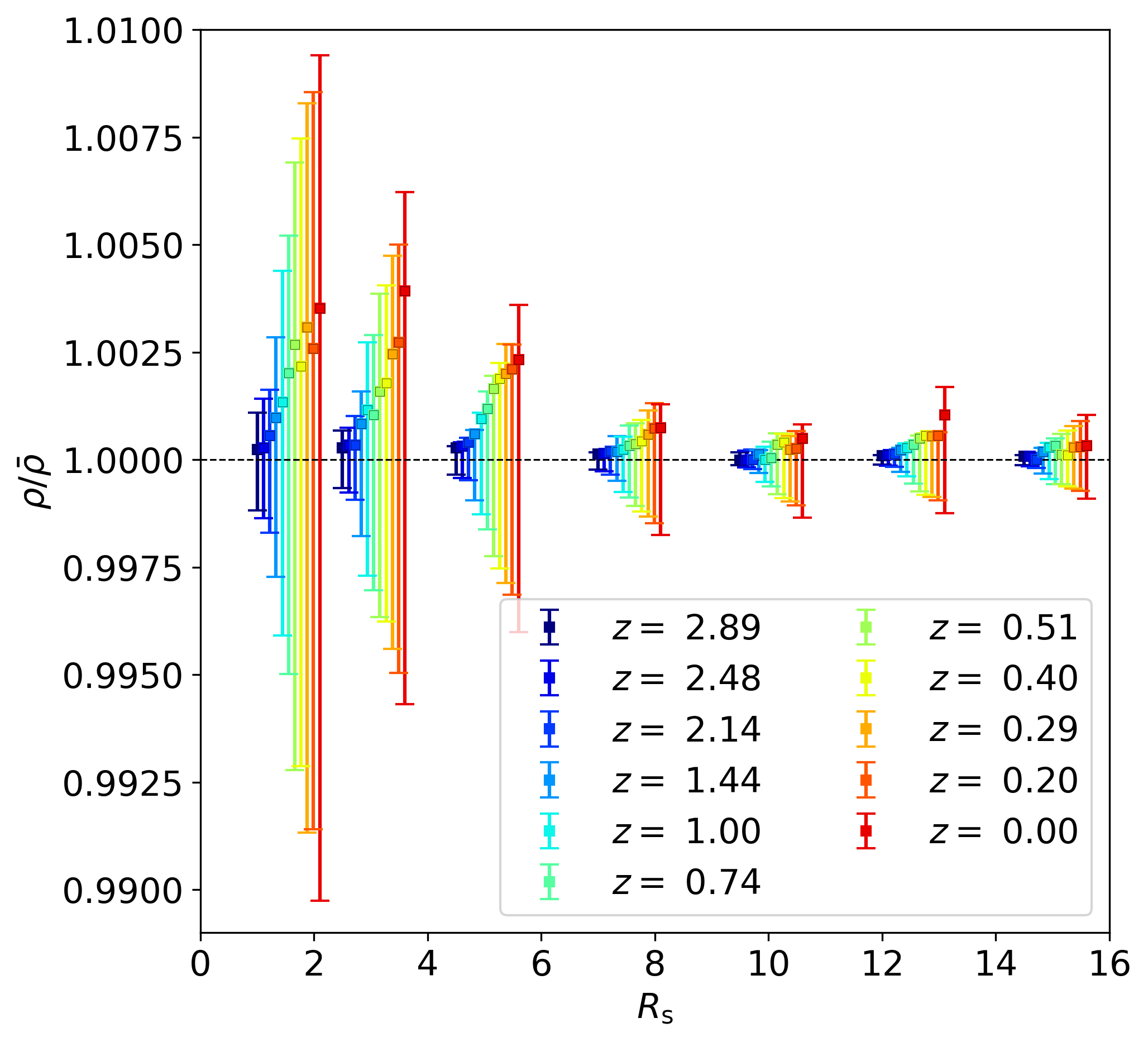}
\caption{Mean density in basins as a function of the smoothing scale $r_s$. The color-code represents the redshift. The scattered points correspond to the median of all basins identified in the SMD velocity field at given $r_s$ and $z$, while the error bars correspond to one standard deviation of the volume weighted density distribution. For the sake of clarity, scattered points of the same smoothing scale are slightly shifted and centered on their corresponding value of $r_s$.}
\label{fig:zmeanrho}
\end{figure}

 Figure \ref{fig:zmeanrho} displays the mean density in basins as a function of the smoothing scale $r_s$. Similarly to Figure \ref{fig:znumber}, the color-code represents the redshift. The scattered points correspond to the median of all basins identified in the SMD velocity field at given $r_s$ and $z$, while the error bars correspond to one standard deviation of the volume weighted density distribution. For greater clarity, scattered points of the same smoothing scale $r_s$ are slightly shifted and centered on their corresponding value of $r_s$. One can observe that, regardless of the redshift and the smoothing scale, basins of attraction have a mean density slightly above 1. The standard deviation of the density field is however strongly redshift dependent - since the density field is more contrasted at lower redshifts the standard deviation  is higher. In otherwords: larger smoothings at low redshifts result in standard deviations of the normalized density field that are equal to that of smaller smoothings at higher redshifts.

\begin{figure}
\includegraphics[width=\linewidth]{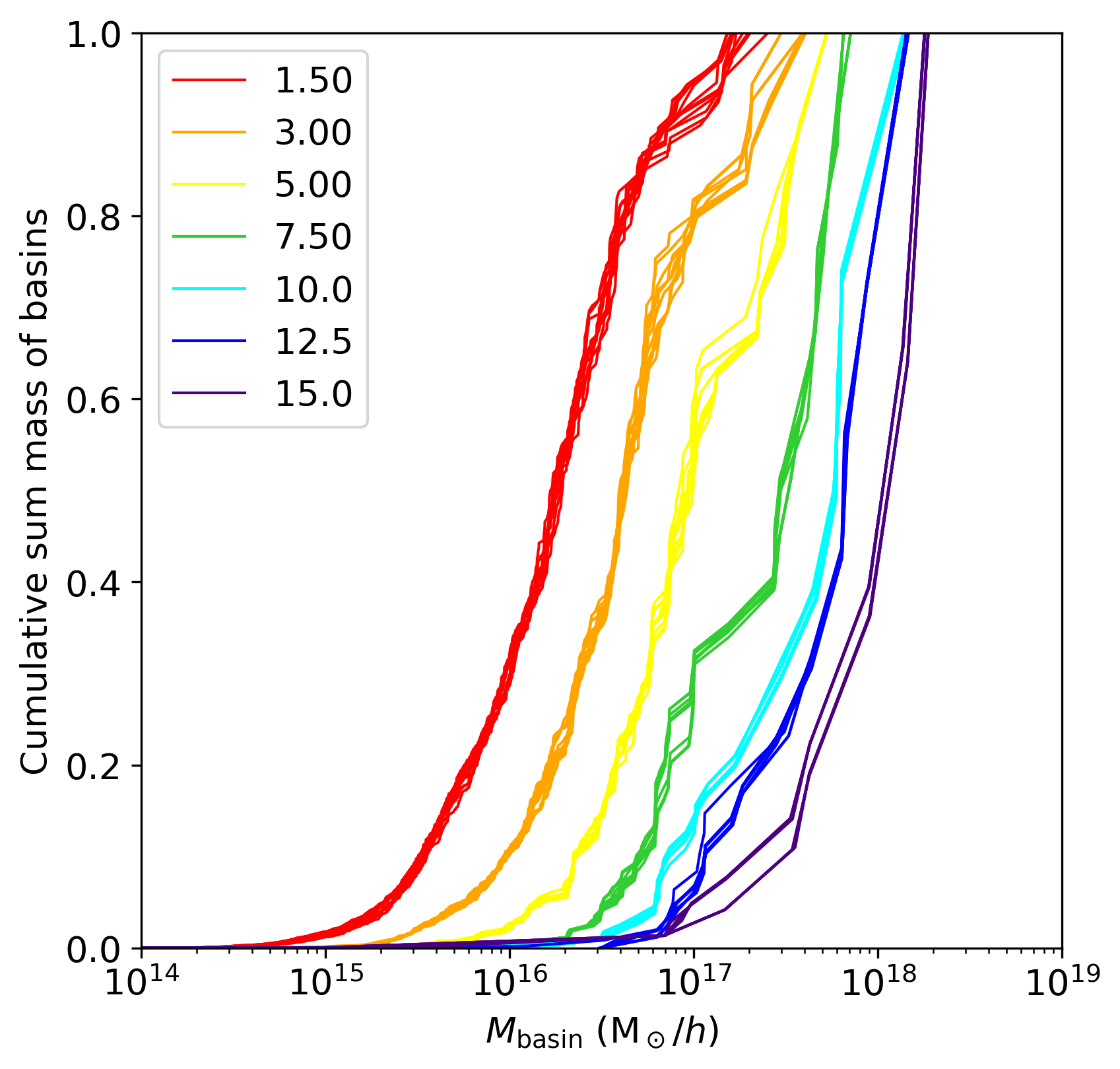}
\caption{Cumulative sum of the mass of basins (normalized by the total mass) as a function of the mass of the basins of attraction segmented. The color code represents the smoothing scale $r_s$ considered. The solid lines of the same color correspond to velocity fields with the same $r_s$ but different redshifts.}
\label{fig:zmass}
\end{figure}

 One can also study the properties of basins, such as their mass and their mass function. Figure \ref{fig:zmass} shows, as a function of the mass of the basins of attraction segmented, the cumulative sum of the mass of basins (normalized by the total mass in the simulation); namely the fraction of the Universe's mass locked up in basins less than a given mass. The color code represents the smoothing scale $r_s$ considered. The solid lines of the same color correspond to velocity fields with the same $r_s$ but at different redshifts. Since these are nearly identical, we haven't displayed them with different line styles. One can see that lines of the same $r_s$ overlap. The cumulative mass function of basins of attraction is independent of the redshift. This is an important result - basins are essentially timeless structures. Again a dependence in the smoothing scale is seen: a large $r_s$ leads to more massive basins of attraction. On the smallest scales half the mass of the universe is in basins smaller than around $10^{16}M_{\odot}$, while on the largest scales half the mass of the Universe is in basins smaller than $10^{18}M_{\odot}$. At a given smoothing we are thus able, for the first time, to quantify whether a basin is ``big'' or ``small''.

\begin{figure}
\includegraphics[width=\linewidth]{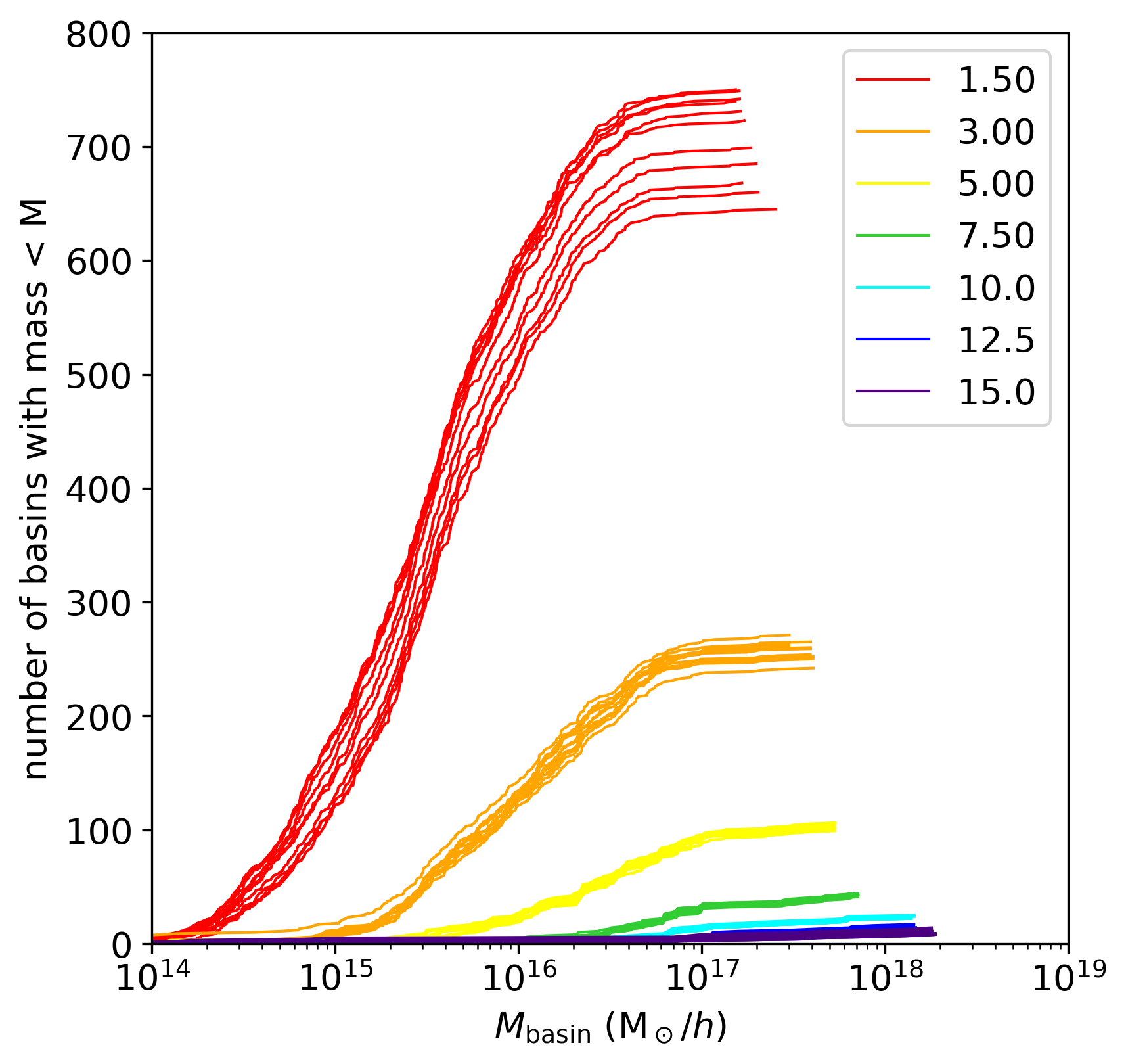}
\caption{Number of basins enclosing a total mass under a given threshold mass $M$ as a function of $M$. Different smoothing scales are considered and represented by different colors (see color code). Solid lines of the same color correspond to velocity fields with the same $r_s$ but different $z$.}
\label{fig:zmass3}
\end{figure}

The number of basins with an enclosed total mass less than a given  mass $M$ is shown in Figure \ref{fig:zmass3} as a function of the value of $M$. The color code is the same as the previous figure. Different colors stand for different smoothing lengths $r_s$ and overlapping solid lines of the same color correspond to the partitioning of velocity fields with the same $r_s$ value but different $z$. The shape of the distribution depends highly on the value of the smoothing length considered: there are fewer total number of basins and these become more massive as $r_s$ increases. However the dependence in redshift is very low as the lines corresponding to the same $r_s$ more or less  overlap, showing that basins have comparable masses from one redshift to another, i.e the mass of basins stay unchanged with time. We do note however that at a given $r_s$ there is some differences, as a function of $z$ for the total basin number.  

\begin{figure}
\includegraphics[width=\linewidth]{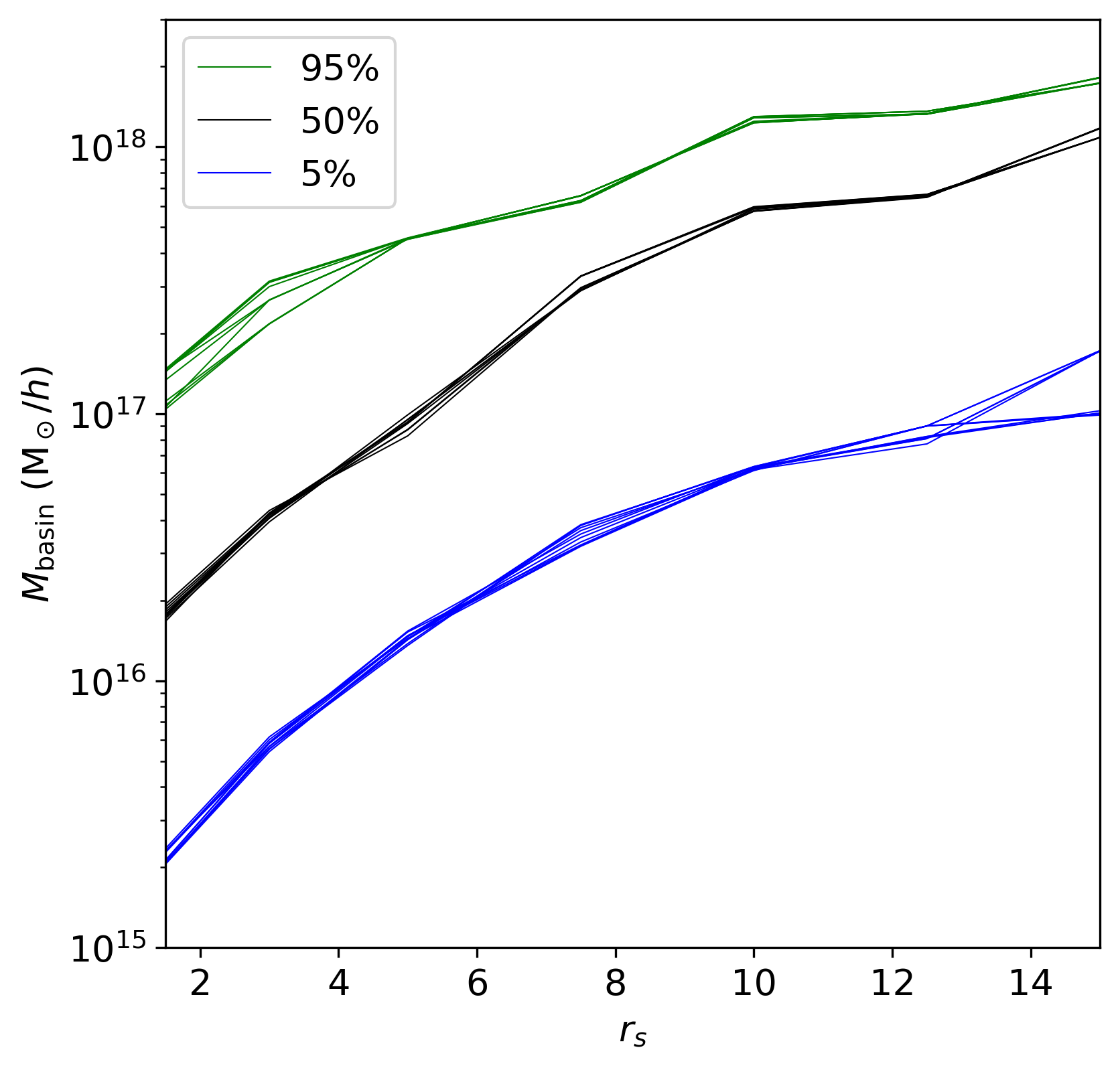}
\caption{Mass of basins of attraction at three levels of the cumulative sum shown in Figure \ref{fig:zmass} is displayed as a function of the smoothing scale $r_s$: 5\% of the total mass in blue, half of the total mass in black and 95\% in green.}
\label{fig:zmass2}
\end{figure}

 Figure \ref{fig:zmass2} shows another way to visualize the dependence of the distribution of basin mass on the smoothing scale. The mass of basins of attraction at three levels of the cumulative sum shown in Figure \ref{fig:zmass2} is displayed as a function of the smoothing scale $r_s$. Here we show the median basin mass and the 90 percentile interval namely, the mass of basins whose sum contains 5 percent of the total (in blue), 50 percent of the the total (in black) and 95 percent (in green).

\section{Conclusion}\label{sec:conclusion}

The methodology considered in this paper and introduced previously in \cite{2019MNRAS.489L...1D} proposes a natural way of defining large scale structures as segmented volumes of universe. These volumes are dynamical gravitational basins computed using the peculiar velocity field of galaxies in cosmological numerical simulations.

This paper studies in detail the methodology of \cite{2019MNRAS.489L...1D} with extensive analysis regarding the nature of basins of attractions in the $\Lambda$CDM Universe. To do so, the Small Multidark simulation ($400$Mpc$/h$ box with $3840^3$ Dark matter particles) has been examined. The various algorithmic parameters involved in the segmentation algortihm have been tested on the simulation's velocity field at $z=0$, as well as the effect of the Gaussian smoothing scale. It has been found, unsurprisingly, that the partitioning of a velocity field depends on the choice of the algortihmic parameters. However, we have identified values of the algorithmic parameters for which the number and the shapes of basins converge, which thus allows a sensible non arbitrary choice of parameters. 

The segmentation is highly dependent with the smoothing scale: as the smoothing scale increases, less basins are identified as they become larger and more massive. This is a physical consequence of smoothing and can be selected depending on the problem at hand.

The cosmic evolution of gravitational basins has also been explored. The properties of basins of attraction have been examined at various redshifts from $z=0$ up to $z=2.89$. Unlike the smoothing scale, no significant dependence in redshift can be seen regarding the number of basins. Even though a larger scatter can been seen at smaller smoothing scales, the number of basins, as well as their mean density and their mass, do not evolve with time. Again, this is expected as basins of attraction are theoretically envisioned to be ``timeless'' entities set by the initial conditions.

The series of tests conducted here considers velocity fields at several redshifts up to 2.89 only, from a $\Lambda$CDM simulation. Additional tests could be pursued at higher redshift, along with applications to simulations of various cosmological models (for example WDM, etc.). Furthermore it is of great interest to see how the nature of a single basin changes if the sampling of the velocity field is degraded. For example, in this paper we have worked with the ``true'' velocity field as determined by a CIC directly from the particle distribution. How would it look if the velocity field was computed from halo velocities? how would it look if the sampling of halo velocities was finer or poorer? If it were homogeneously sampled or if the sampling included observational biases (such as zone of avoidance, magnitude limits, etc). We leave these questions to a future work. 
One could also envisage the possibility to measure cosmological parameters, such as the growth rate of large-scale structure, with the help of the segmentation methodology and gravitational basins.


\section{Acknowledgements}

The authors acknowledge support from the Institut Universitaire de France, the CNES, the Project IDEXLYON at the University of Lyon under the Investments for the Future Program (ANR-16-IDEX-0005). NIL acknowledges acknowledges support from the DFG (DFG-LI 2015/5-1). The authors gratefully acknowledge the Gauss Centre for Supercomputing e.V. (www.gauss-centre.eu) and the Partnership for Advanced Supercomputing in Europe (PRACE, www.prace-ri.eu) for funding the MultiDark simulation project by providing computing time on the GCS Supercomputer SuperMUC at Leibniz Supercomputing Centre (LRZ, www.lrz.de).

\bibliographystyle{mnras}
\bibliography{bibliography}

\begin{thebibliography}{}
\makeatletter
\relax
\def\mn@urlcharsother{\let\do\@makeother \do\$\do\&\do\#\do\^\do\_\do\%\do\~}
\def\mn@doi{\begingroup\mn@urlcharsother \@ifnextchar [ {\mn@doi@}
  {\mn@doi@[]}}
\def\mn@doi@[#1]#2{\def\@tempa{#1}\ifx\@tempa\@empty \href
  {http://dx.doi.org/#2} {doi:#2}\else \href {http://dx.doi.org/#2} {#1}\fi
  \endgroup}
\def\mn@eprint#1#2{\mn@eprint@#1:#2::\@nil}
\def\mn@eprint@arXiv#1{\href {http://arxiv.org/abs/#1} {{\tt arXiv:#1}}}
\def\mn@eprint@dblp#1{\href {http://dblp.uni-trier.de/rec/bibtex/#1.xml}
  {dblp:#1}}
\def\mn@eprint@#1:#2:#3:#4\@nil{\def\@tempa {#1}\def\@tempb {#2}\def\@tempc
  {#3}\ifx \@tempc \@empty \let \@tempc \@tempb \let \@tempb \@tempa \fi \ifx
  \@tempb \@empty \def\@tempb {arXiv}\fi \@ifundefined
  {mn@eprint@\@tempb}{\@tempb:\@tempc}{\expandafter \expandafter \csname
  mn@eprint@\@tempb\endcsname \expandafter{\@tempc}}}

\bibitem[\protect\citeauthoryear{{Alpaslan} et~al.,}{{Alpaslan}
  et~al.}{2014}]{2014MNRAS.438..177A}
{Alpaslan} M.,  et~al., 2014, \mn@doi [\mnras] {10.1093/mnras/stt2136}, \href
  {https://ui.adsabs.harvard.edu/abs/2014MNRAS.438..177A} {438, 177}

\bibitem[\protect\citeauthoryear{{Aragon-Calvo} \& {Yang}}{{Aragon-Calvo} \&
  {Yang}}{2014}]{2014MNRAS.440L..46A}
{Aragon-Calvo} M.~A.,  {Yang} L.~F.,  2014, \mn@doi [\mnras]
  {10.1093/mnrasl/slu009}, \href
  {https://ui.adsabs.harvard.edu/abs/2014MNRAS.440L..46A} {440, L46}

\bibitem[\protect\citeauthoryear{{Arag{\'o}n-Calvo}, {Jones}, {van de Weygaert}
   \& {van der Hulst}}{{Arag{\'o}n-Calvo} et~al.}{2007}]{Aragon-Calvo:2007aa}
{Arag{\'o}n-Calvo} M.~A.,  {Jones} B.~J.~T.,  {van de Weygaert} R.,   {van der
  Hulst} J.~M.,  2007, \mn@doi [\aap] {10.1051/0004-6361:20077880}, \href
  {http://adsabs.harvard.edu/abs/2007A%26A...474..315A} {474, 315}

\bibitem[\protect\citeauthoryear{{Arag{\'o}n-Calvo}, {Platen}, {van de
  Weygaert}  \& {Szalay}}{{Arag{\'o}n-Calvo}
  et~al.}{2010}]{Aragon-Calvo:2010aa}
{Arag{\'o}n-Calvo} M.~A.,  {Platen} E.,  {van de Weygaert} R.,   {Szalay}
  A.~S.,  2010, \mn@doi [\apj] {10.1088/0004-637X/723/1/364}, \href
  {http://adsabs.harvard.edu/abs/2010ApJ...723..364A} {723, 364}

\bibitem[\protect\citeauthoryear{{Bond}, {Kofman}  \& {Pogosyan}}{{Bond}
  et~al.}{1996}]{1996Natur.380..603B}
{Bond} J.~R.,  {Kofman} L.,   {Pogosyan} D.,  1996, \mn@doi [\nat]
  {10.1038/380603a0}, \href
  {https://ui.adsabs.harvard.edu/abs/1996Natur.380..603B} {380, 603}

\bibitem[\protect\citeauthoryear{{Cautun}, {van de Weygaert}  \&
  {Jones}}{{Cautun} et~al.}{2013}]{Cautun:2013aa}
{Cautun} M.,  {van de Weygaert} R.,   {Jones} B.~J.~T.,  2013, \mn@doi [\mnras]
  {10.1093/mnras/sts416}, \href
  {http://adsabs.harvard.edu/abs/2013MNRAS.429.1286C} {429, 1286}

\bibitem[\protect\citeauthoryear{{Courtois}, {Tully}, {Hoffman},
  {Pomar{\`e}de}, {Graziani}  \& {Dupuy}}{{Courtois}
  et~al.}{2017}]{2017ApJ...847L...6C}
{Courtois} H.~M.,  {Tully} R.~B.,  {Hoffman} Y.,  {Pomar{\`e}de} D.,
  {Graziani} R.,   {Dupuy} A.,  2017, \mn@doi [\apjl]
  {10.3847/2041-8213/aa88b2}, \href
  {https://ui.adsabs.harvard.edu/abs/2017ApJ...847L...6C} {847, L6}

\bibitem[\protect\citeauthoryear{{Courtois}, {Kraan-Korteweg}, {Dupuy},
  {Graziani}  \& {Libeskind}}{{Courtois} et~al.}{2019}]{2019MNRAS.490L..57C}
{Courtois} H.~M.,  {Kraan-Korteweg} R.~C.,  {Dupuy} A.,  {Graziani} R.,
  {Libeskind} N.~I.,  2019, \mn@doi [\mnras] {10.1093/mnrasl/slz146}, \href
  {https://ui.adsabs.harvard.edu/abs/2019MNRAS.490L..57C} {490, L57}

\bibitem[\protect\citeauthoryear{{Doroshkevich}}{{Doroshkevich}}{1970}]{1970Ap......6..320D}
{Doroshkevich} A.~G.,  1970, \mn@doi [Astrophysics] {10.1007/BF01001625}, \href
  {https://ui.adsabs.harvard.edu/abs/1970Ap......6..320D} {6, 320}

\bibitem[\protect\citeauthoryear{{Dupuy} et~al.,}{{Dupuy}
  et~al.}{2019}]{2019MNRAS.489L...1D}
{Dupuy} A.,  et~al., 2019, \mn@doi [\mnras] {10.1093/mnrasl/slz115}, \href
  {https://ui.adsabs.harvard.edu/abs/2019MNRAS.489L...1D} {489, L1}

\bibitem[\protect\citeauthoryear{{Falck} \& {Neyrinck}}{{Falck} \&
  {Neyrinck}}{2015}]{2015MNRAS.450.3239F}
{Falck} B.,  {Neyrinck} M.~C.,  2015, \mn@doi [\mnras] {10.1093/mnras/stv879},
  \href {https://ui.adsabs.harvard.edu/abs/2015MNRAS.450.3239F} {450, 3239}

\bibitem[\protect\citeauthoryear{{Falck}, {Neyrinck}  \& {Szalay}}{{Falck}
  et~al.}{2012}]{2012ApJ...754..126F}
{Falck} B.~L.,  {Neyrinck} M.~C.,   {Szalay} A.~S.,  2012, \mn@doi [\apj]
  {10.1088/0004-637X/754/2/126}, \href
  {https://ui.adsabs.harvard.edu/abs/2012ApJ...754..126F} {754, 126}

\bibitem[\protect\citeauthoryear{{Forero-Romero}, {Hoffman}, {Gottl{\"o}ber},
  {Klypin}  \& {Yepes}}{{Forero-Romero} et~al.}{2009}]{Forero-Romero:2009aa}
{Forero-Romero} J.~E.,  {Hoffman} Y.,  {Gottl{\"o}ber} S.,  {Klypin} A.,
  {Yepes} G.,  2009, \mn@doi [\mnras] {10.1111/j.1365-2966.2009.14885.x}, \href
  {http://adsabs.harvard.edu/abs/2009MNRAS.396.1815F} {396, 1815}

\bibitem[\protect\citeauthoryear{{Geller} \& {Huchra}}{{Geller} \&
  {Huchra}}{1989}]{1989Sci...246..897G}
{Geller} M.~J.,  {Huchra} J.~P.,  1989, \mn@doi [Science]
  {10.1126/science.246.4932.897}, \href
  {https://ui.adsabs.harvard.edu/abs/1989Sci...246..897G} {246, 897}

\bibitem[\protect\citeauthoryear{{Giovanelli} \& {Haynes}}{{Giovanelli} \&
  {Haynes}}{1985}]{1985AJ.....90.2445G}
{Giovanelli} R.,  {Haynes} M.~P.,  1985, \mn@doi [\aj] {10.1086/113949}, \href
  {https://ui.adsabs.harvard.edu/abs/1985AJ.....90.2445G} {90, 2445}

\bibitem[\protect\citeauthoryear{{Gonz{\'a}lez} \& {Padilla}}{{Gonz{\'a}lez} \&
  {Padilla}}{2010}]{2010MNRAS.407.1449G}
{Gonz{\'a}lez} R.~E.,  {Padilla} N.~D.,  2010, \mn@doi [\mnras]
  {10.1111/j.1365-2966.2010.17015.x}, \href
  {https://ui.adsabs.harvard.edu/abs/2010MNRAS.407.1449G} {407, 1449}

\bibitem[\protect\citeauthoryear{{Gott}, {Juri{\'c}}, {Schlegel}, {Hoyle},
  {Vogeley}, {Tegmark}, {Bahcall}  \& {Brinkmann}}{{Gott}
  et~al.}{2005}]{2005ApJ...624..463G}
{Gott} J.~Richard I.,  {Juri{\'c}} M.,  {Schlegel} D.,  {Hoyle} F.,  {Vogeley}
  M.,  {Tegmark} M.,  {Bahcall} N.,   {Brinkmann} J.,  2005, \mn@doi [\apj]
  {10.1086/428890}, \href
  {https://ui.adsabs.harvard.edu/abs/2005ApJ...624..463G} {624, 463}

\bibitem[\protect\citeauthoryear{{Graziani}, {Courtois}, {Lavaux}, {Hoffman},
  {Tully}, {Copin}  \& {Pomar{\`e}de}}{{Graziani}
  et~al.}{2019}]{2019MNRAS.488.5438G}
{Graziani} R.,  {Courtois} H.~M.,  {Lavaux} G.,  {Hoffman} Y.,  {Tully} R.~B.,
  {Copin} Y.,   {Pomar{\`e}de} D.,  2019, \mn@doi [\mnras]
  {10.1093/mnras/stz078}, \href
  {https://ui.adsabs.harvard.edu/abs/2019MNRAS.488.5438G} {488, 5438}

\bibitem[\protect\citeauthoryear{{Hahn}, {Porciani}, {Carollo}  \&
  {Dekel}}{{Hahn} et~al.}{2007}]{Hahn:2007aa}
{Hahn} O.,  {Porciani} C.,  {Carollo} C.~M.,   {Dekel} A.,  2007, \mn@doi
  [\mnras] {10.1111/j.1365-2966.2006.11318.x}, \href
  {http://adsabs.harvard.edu/abs/2007MNRAS.375..489H} {375, 489}

\bibitem[\protect\citeauthoryear{{Hoffman}, {Metuki}, {Yepes}, {Gottl{\"o}ber},
  {Forero-Romero}, {Libeskind}  \& {Knebe}}{{Hoffman}
  et~al.}{2012}]{Hoffman:2012aa}
{Hoffman} Y.,  {Metuki} O.,  {Yepes} G.,  {Gottl{\"o}ber} S.,  {Forero-Romero}
  J.~E.,  {Libeskind} N.~I.,   {Knebe} A.,  2012, \mn@doi [\mnras]
  {10.1111/j.1365-2966.2012.21553.x}, \href
  {http://adsabs.harvard.edu/abs/2012MNRAS.425.2049H} {425, 2049}

\bibitem[\protect\citeauthoryear{{Hoffman}, {Pomar{\`e}de}, {Tully}  \&
  {Courtois}}{{Hoffman} et~al.}{2017}]{2017NatAs...1E..36H}
{Hoffman} Y.,  {Pomar{\`e}de} D.,  {Tully} R.~B.,   {Courtois} H.~M.,  2017,
  \mn@doi [Nature Astronomy] {10.1038/s41550-016-0036}, \href
  {https://ui.adsabs.harvard.edu/abs/2017NatAs...1E..36H} {1, 0036}

\bibitem[\protect\citeauthoryear{{Kitaura} \& {Angulo}}{{Kitaura} \&
  {Angulo}}{2012}]{2012MNRAS.425.2443K}
{Kitaura} F.-S.,  {Angulo} R.~E.,  2012, \mn@doi [\mnras]
  {10.1111/j.1365-2966.2012.21614.x}, \href
  {https://ui.adsabs.harvard.edu/abs/2012MNRAS.425.2443K} {425, 2443}

\bibitem[\protect\citeauthoryear{{Klypin}, {Yepes}, {Gottl{\"o}ber}, {Prada}
  \& {He{\ss}}}{{Klypin} et~al.}{2016}]{2016MNRAS.457.4340K}
{Klypin} A.,  {Yepes} G.,  {Gottl{\"o}ber} S.,  {Prada} F.,   {He{\ss}} S.,
  2016, \mn@doi [\mnras] {10.1093/mnras/stw248}, \href
  {https://ui.adsabs.harvard.edu/abs/2016MNRAS.457.4340K} {457, 4340}

\bibitem[\protect\citeauthoryear{{Kraan-Korteweg}, {Cluver}, {Bilicki},
  {Jarrett}, {Colless}, {Elagali}, {B{\"o}hringer}  \& {Chon}}{{Kraan-Korteweg}
  et~al.}{2017}]{2017MNRAS.466L..29K}
{Kraan-Korteweg} R.~C.,  {Cluver} M.~E.,  {Bilicki} M.,  {Jarrett} T.~H.,
  {Colless} M.,  {Elagali} A.,  {B{\"o}hringer} H.,   {Chon} G.,  2017, \mn@doi
  [\mnras] {10.1093/mnrasl/slw229}, \href
  {https://ui.adsabs.harvard.edu/abs/2017MNRAS.466L..29K} {466, L29}

\bibitem[\protect\citeauthoryear{{Leclercq}, {Jasche}, {Lavaux}, {Wandelt}  \&
  {Percival}}{{Leclercq} et~al.}{2017}]{Leclercq:2017aa}
{Leclercq} F.,  {Jasche} J.,  {Lavaux} G.,  {Wandelt} B.,   {Percival} W.,
  2017, \mn@doi [\jcap] {10.1088/1475-7516/2017/06/049}, \href
  {http://adsabs.harvard.edu/abs/2017JCAP...06..049L} {6, 049}

\bibitem[\protect\citeauthoryear{{Libeskind}, {Hoffman}  \&
  {Gottl{\"o}ber}}{{Libeskind} et~al.}{2014}]{2014MNRAS.441.1974L}
{Libeskind} N.~I.,  {Hoffman} Y.,   {Gottl{\"o}ber} S.,  2014, \mn@doi [\mnras]
  {10.1093/mnras/stu629}, \href
  {https://ui.adsabs.harvard.edu/abs/2014MNRAS.441.1974L} {441, 1974}

\bibitem[\protect\citeauthoryear{{Libeskind} et~al.,}{{Libeskind}
  et~al.}{2018}]{Libeskind:2018aa}
{Libeskind} N.~I.,  et~al., 2018, \mn@doi [\mnras] {10.1093/mnras/stx1976},
  \href {http://adsabs.harvard.edu/abs/2018MNRAS.473.1195L} {473, 1195}

\bibitem[\protect\citeauthoryear{{Lietzen} et~al.,}{{Lietzen}
  et~al.}{2016}]{2016A&A...588L...4L}
{Lietzen} H.,  et~al., 2016, \mn@doi [\aap] {10.1051/0004-6361/201628261},
  \href {https://ui.adsabs.harvard.edu/abs/2016A&A...588L...4L} {588, L4}

\bibitem[\protect\citeauthoryear{{Lynden-Bell}, {Faber}, {Burstein}, {Davies},
  {Dressler}, {Terlevich}  \& {Wegner}}{{Lynden-Bell}
  et~al.}{1988}]{1988ApJ...326...19L}
{Lynden-Bell} D.,  {Faber} S.~M.,  {Burstein} D.,  {Davies} R.~L.,  {Dressler}
  A.,  {Terlevich} R.~J.,   {Wegner} G.,  1988, \mn@doi [\apj]
  {10.1086/166066}, \href
  {https://ui.adsabs.harvard.edu/abs/1988ApJ...326...19L} {326, 19}

\bibitem[\protect\citeauthoryear{{Platen}, {van de Weygaert}  \&
  {Jones}}{{Platen} et~al.}{2007}]{2007MNRAS.380..551P}
{Platen} E.,  {van de Weygaert} R.,   {Jones} B. J.~T.,  2007, \mn@doi [\mnras]
  {10.1111/j.1365-2966.2007.12125.x}, \href
  {https://ui.adsabs.harvard.edu/abs/2007MNRAS.380..551P} {380, 551}

\bibitem[\protect\citeauthoryear{{Pomar{\`e}de}, {Tully}, {Hoffman}  \&
  {Courtois}}{{Pomar{\`e}de} et~al.}{2015}]{2015ApJ...812...17P}
{Pomar{\`e}de} D.,  {Tully} R.~B.,  {Hoffman} Y.,   {Courtois} H.~M.,  2015,
  \mn@doi [\apj] {10.1088/0004-637X/812/1/17}, \href
  {https://ui.adsabs.harvard.edu/abs/2015ApJ...812...17P} {812, 17}

\bibitem[\protect\citeauthoryear{{Ramachandra} \& {Shandarin}}{{Ramachandra} \&
  {Shandarin}}{2015}]{2015MNRAS.452.1643R}
{Ramachandra} N.~S.,  {Shandarin} S.~F.,  2015, \mn@doi [\mnras]
  {10.1093/mnras/stv1389}, \href
  {https://ui.adsabs.harvard.edu/abs/2015MNRAS.452.1643R} {452, 1643}

\bibitem[\protect\citeauthoryear{{Rodr{\'{\i}}guez-Puebla}, {Behroozi},
  {Primack}, {Klypin}, {Lee}  \& {Hellinger}}{{Rodr{\'{\i}}guez-Puebla}
  et~al.}{2016}]{2016MNRAS.462..893R}
{Rodr{\'{\i}}guez-Puebla} A.,  {Behroozi} P.,  {Primack} J.,  {Klypin} A.,
  {Lee} C.,   {Hellinger} D.,  2016, \mn@doi [\mnras] {10.1093/mnras/stw1705},
  \href {http://adsabs.harvard.edu/abs/2016MNRAS.462..893R} {462, 893}

\bibitem[\protect\citeauthoryear{{Sousbie}}{{Sousbie}}{2011}]{Sousbie:2011aa}
{Sousbie} T.,  2011, \mn@doi [\mnras] {10.1111/j.1365-2966.2011.18394.x}, \href
  {http://adsabs.harvard.edu/abs/2011MNRAS.414..350S} {414, 350}

\bibitem[\protect\citeauthoryear{{Sousbie}, {Pichon}, {Courtois}, {Colombi}  \&
  {Novikov}}{{Sousbie} et~al.}{2008}]{Sousbie:2008aa}
{Sousbie} T.,  {Pichon} C.,  {Courtois} H.,  {Colombi} S.,   {Novikov} D.,
  2008, \mn@doi [\apjl] {10.1086/523669}, \href
  {http://adsabs.harvard.edu/abs/2008ApJ...672L...1S} {672, L1}

\bibitem[\protect\citeauthoryear{{Tempel}, {Stoica}, {Mart{\'\i}nez},
  {Liivam{\"a}gi}, {Castellan}  \& {Saar}}{{Tempel}
  et~al.}{2014}]{2014MNRAS.438.3465T}
{Tempel} E.,  {Stoica} R.~S.,  {Mart{\'\i}nez} V.~J.,  {Liivam{\"a}gi} L.~J.,
  {Castellan} G.,   {Saar} E.,  2014, \mn@doi [\mnras] {10.1093/mnras/stt2454},
  \href {https://ui.adsabs.harvard.edu/abs/2014MNRAS.438.3465T} {438, 3465}

\bibitem[\protect\citeauthoryear{{Tempel}, {Stoica}, {Kipper}  \&
  {Saar}}{{Tempel} et~al.}{2016}]{2016A&C....16...17T}
{Tempel} E.,  {Stoica} R.~S.,  {Kipper} R.,   {Saar} E.,  2016, \mn@doi
  [Astronomy and Computing] {10.1016/j.ascom.2016.03.004}, \href
  {https://ui.adsabs.harvard.edu/abs/2016A&C....16...17T} {16, 17}

\bibitem[\protect\citeauthoryear{{Tully} \& {Fisher}}{{Tully} \&
  {Fisher}}{1987}]{1987ang..book.....T}
{Tully} R.~B.,  {Fisher} J.~R.,  1987, {Atlas of Nearby Galaxies}

\bibitem[\protect\citeauthoryear{{Tully}, {Courtois}, {Hoffman}  \&
  {Pomar{\`e}de}}{{Tully} et~al.}{2014}]{Tully:2014aa}
{Tully} R.~B.,  {Courtois} H.,  {Hoffman} Y.,   {Pomar{\`e}de} D.,  2014,
  \mn@doi [\nat] {10.1038/nature13674}, \href
  {http://adsabs.harvard.edu/abs/2014Natur.513...71T} {513, 71}

\makeatother
\end{thebibliography}

\bsp	
\label{lastpage}

\end{document}